\def\year{2022}\relax
\documentclass[letterpaper]{article} 
\usepackage{aaai22}  
\usepackage{times}  
\usepackage{helvet}  
\usepackage{courier}  
\usepackage[hyphens]{url}  
\usepackage{graphicx} 
\usepackage{natbib}  
\usepackage{caption} 
\DeclareCaptionStyle{ruled}{labelfont=normalfont,labelsep=colon,strut=off} 
\usepackage{algorithm}

\usepackage{adjustbox}
\usepackage{multirow}
\usepackage{multicol}
\usepackage{booktabs,colortbl}
\usepackage{amsmath}
\usepackage{algpseudocode}
\usepackage{subcaption}
\usepackage{blindtext}
\usepackage{dblfloatfix}
\usepackage[dvipsnames, x11names]{xcolor}
\usepackage[explicit]{titlesec}
\usepackage{lipsum}
\usepackage{tikz}
\usepackage{amsmath,amssymb}
\usepackage{xcolor}

\usepackage{newfloat}
\usepackage{listings}
\floatstyle{ruled}
\newfloat{listing}{tb}{lst}{}
\floatname{listing}{Listing}

\title{Effects of Research Paper Promotion via ArXiv and X}
\author {
    Chhandak Bagchi\textsuperscript{\rm 1},
    Eric Malmi\thanks{Now at Google, Z{\"u}rich.} \textsuperscript{\rm 2}, 
    Przemyslaw Grabowicz \textsuperscript{\rm 1}
}
\affiliations {
    \textsuperscript{\rm 1} Manning College of Information and Computer Sciences, University of Massachusetts Amherst\\
    \textsuperscript{\rm 2} Aalto University\\
    
    cbagchi@cs.umass.edu, eric.malmi@gmail.com, grabowicz@cs.umass.edu
}

\begin{document}

\maketitle

\begin{abstract}

In the evolving landscape of scientific publishing, it is important to understand the drivers of high-impact research, to equip scientists with actionable strategies to enhance the reach of their work, and to understand trends in the use of modern scientific publishing tools to inform their further development. Here, based on a dataset of over 0.5 million publications in computer science and physics, we study trends in the use of early preprint publications and revisions on ArXiv and the use of X (formerly Twitter) for promotion of such papers. We find that early submissions to ArXiv and promotion on X have soared in recent years. Estimating the effect that the use of each of these modern affordances has on the number of citations of scientific publications, we find that peer-reviewed conference papers in computer science that are submitted early to ArXiv gain on average $21.1 \pm 17.4$ more citations, revised on ArXiv gain $18.4 \pm 17.6$ more citations, and promoted on X gain $44.4 \pm 8$ more citations in the first 5 years from an initial publication. In contrast, journal articles in physics experience comparatively lower boosts in citation counts, with increases of $3.9 \pm 1.1$, $4.3 \pm 0.9$, and $6.9 \pm 3.5$ citations respectively for the same interventions. Our results show that promoting one's work on ArXiv or X has a large impact on the number of citations, as well as the number of influential citations computed by Semantic Scholar, and thereby on the career of researchers. These effects are present also for publications in physics, but they are relatively smaller. The larger relative effect sizes, effects of promotion acumulating over time, and elevated unpredictability of the number of citations in computer science than in physics suggest a greater role of world-of-mouth spreading in computer science than in physics. We discuss the far-reaching implications of these findings for future scientific publishing systems and measures of scientific impact.

\end{abstract}

\section{Introduction}

Scientific publications started in the $17^{th}$ century when decisions about whether a paper would be published were made by a select group of experts. As scientific publications proliferated in the $20^{th}$ century, editors of these journals alone could no longer make informed decisions about publications for such a large volume of papers. This led to peer review becoming a standard practice in academia. In the 1990s, the advent of the Web greatly helped with the dissemination of scientific publications. 
Modern science offers affordances such as publishing preprints before peer-review, revising preprints, promoting scientific works on social media, and viewing researcher profiles listing all co-authored publications and the overall number of citations. 
However, it is not clear whether the use of these affordances is on the rise and what is their impact.

Open-access e-print repositories such as ArXiv, which was founded by a physicist, and open publishing platforms such as OpenReview.net, founded by a computer scientist, are at the forefront of this change~\cite{sutton_popularity_2017}, facilitating swift exchange of scientific information. 
ArXiv hosts 2 million e-prints in various technical fields, including computer science and physics.
These platforms enable early publications and later revisions of unreviewed manuscripts. 
Then, once a preprint is published, the work can be instantly promoted on social media. 
Early preprint publication and promotion on social media likely contribute to the number of citations of respective papers, which arguably is the most popular measure of scientific impact. This phenomenon incentivizes scientists to publish early preprints and promote them on social media, without proper peer review of their works, suggesting that the publication of preprints and promotion on social media will become standard practices, if not the case already. 

Unfortunately, these modern affordances of scientific publishing process, can lead to the spread of questionable scientific results. 
An instance of this involves COVID-19 studies published on preprint repositories prior to peer-review that were later retracted due to ethical concerns or research misconduct. These publications still managed to garner up to 593 citations, thereby disseminating misinformation \cite{syed2023covid}. Many of these studies circulated on platforms such as Reddit and 4chan, where their findings were amplified, and premature results were frequently misinterpreted to make future predictions \cite{yudhoatmojo2023understanding}. Additionally, prior works suggest that social influence make it more challenging to distinguish between high and low quality content~\cite{salganik2006experimental}, including in the context of scientific publications~\cite{fisher2015social, baddeley2015herding, resnik2020bias}, and recent research introducing novel measures of scientific impact discusses a correction for such social impact~\cite{ke2023network}.

In this context, it is not surprising that not all venues embrace early unreviewed scientific manuscript publications. Some major computer science conferences, 
enforce anonymity periods on ArXiv submissions before conference submission to uphold double-blind review policies. 
Their critics argue this might decrease paper visibility and subsequent citations. 

\begin{figure}[t]
\includegraphics[width=0.99\columnwidth]{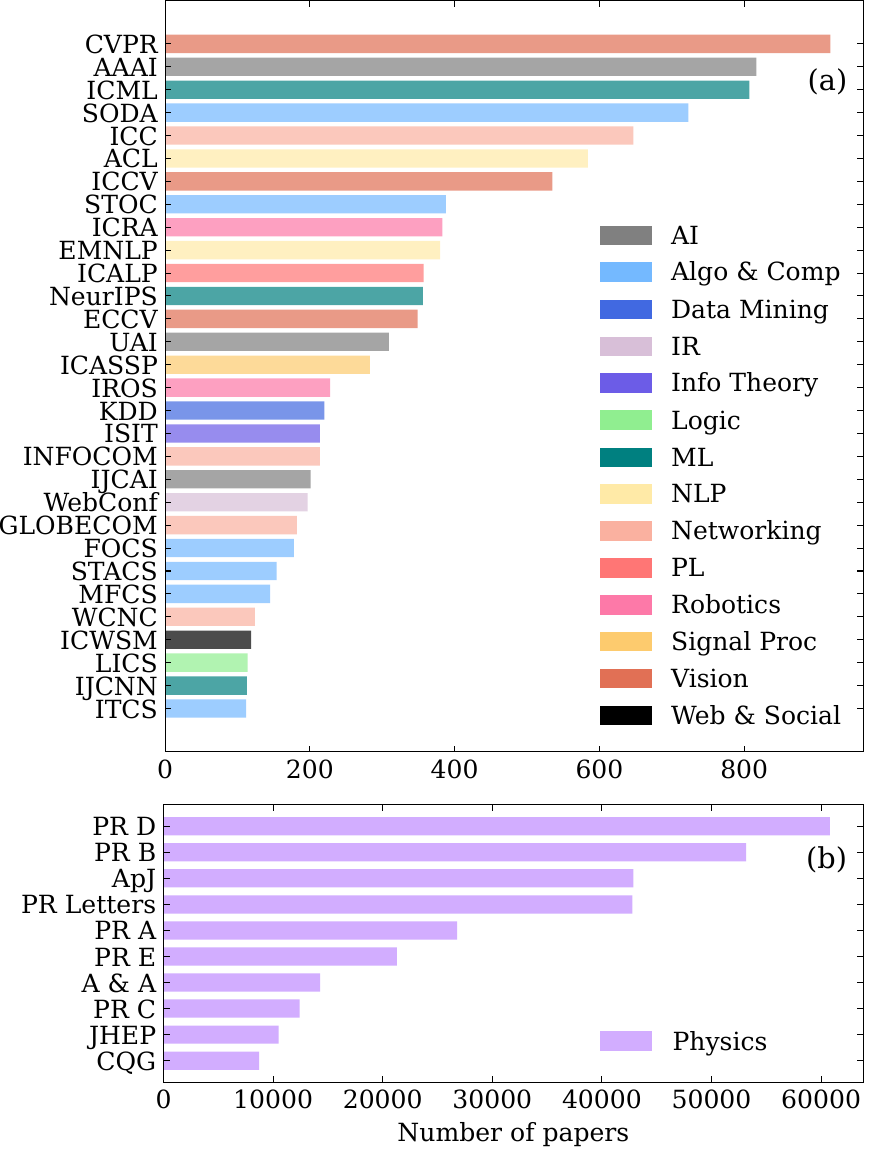}
\caption{(a) Top-30 computer science conferences by paper count in our cross-referenced dataset. The colors correspond to research areas in computer science according to~\citet{sutton_popularity_2017}. (b) Top-10 physics journals by paper count. ``PR'' refers to Physical Review and ``CQG'' referes to Classical and Quantum Gravity.}
\label{fig:papers_in_top_conf}
\end{figure}


Despite these potential implications of early preprint publications and their promotion on social media, there is a lack of comprehensive, large-scale studies on the use of these affordances and their impact on the number of citations. 
This study bridges this gap and reflects on the future of scientific publishing. 
The study focuses on the fields of computer science and physics due to their rapid adoption of new scientific publishing technologies. 
We address the following research questions.

\vspace{2pt}
\textit{\textbf{RQ1}: Are early submissions and revisions on ArXiv, and the practice of tweeting about research publications, emerging as the de facto standards?}
\vspace{2pt}


ArXiv offers next-day publication that can lead to a first-mover advantage and boasts of over 5 million monthly active users~\cite{tech2023arxiv}. 
Publishing first on ArXiv, rather than in a peer-reviewed venue, may impact the number of citations.

\vspace{2pt}
\textit{\textbf{RQ2}: What is the effect of publishing a paper on ArXiv before publishing it in conference proceedings on the number of citations of such a paper?}
\vspace{2pt}



ArXiv also offers the possibility of revising one's work.
On the one hand, a revision may lead to a better publication, especially after receiving expert feedback. On the other hand, if the paper was not fully developed at the time of preprint publication, this could negatively affect its perception among scientists and the number of citations it will receive.  

\vspace{2pt}
\textit{\textbf{RQ3}: What is the effect of revising a paper on ArXiv on the number of citations of such a paper?}
\vspace{2pt}



Next, we investigate the effect of dissemination of scientific research on X.

\vspace{2pt}
\textit{\textbf{RQ4.1}: Can promoting a paper on X (formerly known as Twitter) result in more citations of the paper?}

\textit{\textbf{RQ4.2}: Is the effect of tweets by paper authors different than that of tweets by other users?}
\vspace{2pt}



Next, we hypothesize that more modern measures of scientific impact may be less susceptible to influence of social media, following the intuition that important citations correspond to a deep semantic connection between the citing and cited papers.

\vspace{2pt}
\textit{\textbf{RQ5}: Do newer methods for judging article-level research impact, particularly Semantic Scholar's highly influential citations, show reduced sensitivity to social exposures on X in comparison to the traditional number of citations?}
\vspace{2pt}

Finally, we study the differences in treatment effects between publications in computer science and physics, while aiming to understand the underlying mechanisms that might contribute to these differences.

\vspace{2pt}
\textit{\textbf{RQ6}: How are the effects of social promotion different for computer science compared to physics?}
\vspace{2pt}


These research questions have not been sufficiently addressed yet, since it is challenging to answer them quantitatively. Performing a controlled experiment across multiple venues over multiple years through intervention is intractable. Calculating treatment effects using observational data is difficult because it requires bridging mutliple datasets. 
To this end, we need information about peer-reviewed publications, ArXiv publications, relevant posts on X, and \textit{fine-grained} information about the dates of publications and citations. Another challenge is that there are multiple factors that could confound treatment effect estimates. 

In this paper, we investigate publications in computer science and physics. While computer science publications predominantly occur through conference proceedings, physics publications primarily take the form of journal articles. Both fields contribute significantly to ArXiv.
However, the submission rate of computer science papers to ArXiv has drastically increased at a rate 3.5 times higher than that of physics over the last decade~\cite{ArxivUsageStats}. 
To answer these research questions, we cross-reference six data sources: Semantic Scholar, ArXiv, WikiCFP, Crossref, Altmetric.com, and X. The resulting dataset includes the largest and most prestigious computer science  (Fig.~\ref{fig:papers_in_top_conf}(a)) and physics (Fig. \ref{fig:papers_in_top_conf}(b)) venues, such as ICML and PRL, respectively. The dataset includes 18,113 computer science and 501,766 physics papers spanning 34 years. 
We release this dataset publicly, including author and citation information, publication dates, and ArXiv submission data.\footnote{Code and dataset release: http://bit.ly/48OSnE1}
We define treatments that address our four main research questions (RQ2-5) and study their effects on the number of citations of a paper (RQ2-4), or influential citations (RQ5), in the 5 years since the date of first publication (on ArXiv or conference proceedings). We use the terms ArXiv-first effect (RQ2), revision effect (RQ3), effect of tweeting (RQ4.1), and effect of author(s) tweeting (RQ4.2) to refer to these effects in the remainder of this paper. 
To estimate treatment effects, we rely on state-of-the-art doubly robust estimation. 
We estimate the effects while controlling for potential confounders, such as \textit{longer time to accrue citations}
and \textit{self-selection}
of better papers in hot research areas to be posted on ArXiv before publishing in a peer-reviewed venue. 
Finally, we compare the measured effects across publications in computer science and physics (RQ6).

We recognize the limitations of using citations as a sole metric for assessing scientific impact. For example, papers with erroneous findings can still accumulate numerous citations as researchers cite them to correct the record. Our study demonstrates that citations are not only susceptible to such inherent flaws, but also highly influenced by social promotion. Our results have implications for future publishing models and metrics for gauging scientific impact, as we discuss in the last section of this manuscript, which are particularly relevant given that the computer science and physics communities are uniquely positioned to lead the development of novel publishing models (as exemplified by ArXiv and OpenReview.net) and scientific impact measures~\cite{ke2023network}.

\section{Related Work}

While extensive literature has studied characteristics of open access publications, described next, in this study we investigate a distinct set of questions. 
Our inquiry centers on the impact of submitting earlier to ArXiv than to a peer-reviewed venue, and subsequently revising. Additionally, we investigate the effect of promoting research on social media. While \citet{feldman_citation_2018} have explored the ArXiv-first effect (RQ2) in a smaller study without doubly robust estimation, the effects outlined in RQ1 and RQ3-6 remain unexplored or underexplored.


\subsection{Trends in scientific publishing (RQ1)} 

Publishing on ArXiv and promoting a paper on X have increased in popularity in the computer science community. 
\citet{sutton_popularity_2017} observed a remarkable surge in ArXiv publishing within computer science, reporting that 23\% of all papers in 2017 were published on ArXiv (among which 56\% were preprints), compared to just 1\% in 2007. 
We study trends in early ArXiv submissions, revisions, and social media promotion for computer science and physics publications from 2013 to 2022. 

\subsection{Publishing preprints on ArXiv (RQ2) and the open access effect} 

We measure the effects of submitting a version of one's paper to ArXiv before the conference, i.e., the ArXiv-first effect. \citet{elazar2023estimating} also studied early ArXiving but focused on its effect on conference acceptance rather than citations which is our focus. \citet{feldman_citation_2018}, on the other hand, performed an analysis on citations and found that there is a 65\% increase in citations that can be attributed to the ArXiv-first effect. This study, however, analyzed only the top 16 conferences in computer science, used a subset of the control variables we take into account, and did not apply doubly robust estimation.




Multiple prior works find that open access (OA) papers have higher number of citations~\cite{harnad2004comparing, antelman2004open, fu_releasing_2019, alkhawtani_citation_2020}. This relation was thought to be causal until a study by~\citet{kurtz_effect_2005} about research articles in astrophysics on ArXiv showed that this difference in the number of citations can be explained by two sources of bias that were previously unaccounted for. First, the \textit{early view} bias refers to the situation where articles that are submitted early to a preprint repository, before peer-review publications, have more time to accrue citations. Second, the \textit{self-selection} bias refers to the phenomenon that better papers are selected to be published in preprint repositories in addition to peer-reviewed venues.
\citet{moed2007effect} states that the self-selection bias can occur due to two reasons: i) the author is very well-reputed and is confident about putting out unfinished work (established author bias), ii) the author, irrespective of their reputation in the field, publish their better works on ArXiv (quality bias). These papers in some cases can be unfinished and the authors keep revising them with time. Controlling for these two biases makes the OA advantage negligible for papers in astrophysics.
Since then, there have been multiple studies in different disciplines that study these effects.~\citet*{davis_does_2007} find that for mathematics papers there is no significant early view effect and the difference in citations between OA and non-OA can be explained by the self-selection bias. More specifically, they find that there is a quality differential between papers that get submitted to ArXiv compared to papers that do not.

While we do not directly compute the OA effect (comparing the impact of submitting to ArXiv against not doing so) in this paper, we leverage insights from these studies to inform our work. We control for the early view bias by only considering citations over a fixed 5-year window after the date of first publication, independently whether it is on ArXiv or in conference proceedings. We also control for self-selection bias using a combination of author characteristics and conference characteristics
to lessen the influence of established author and quality biases. 


\subsection{Revising preprints (RQ3)}
We explore the impact of paper revisions on citations (termed the "revision effect"). The prior research on this topic is scarce. \citet{rigby_journal_2018} examined the correlation between revisions and the number of citations for a single social sciences journal. We perform a study at a much larger scale for computer science and physics publications and use treatment effect estimation techniques to account for confounders.

\subsection{Dissemination on social media (RQ4)}
The internet has profoundly changed how scientists share their research, with platforms like X now serving as essential tools for dissemination of scientific information. According to \citet{wang_preprints_2020}, preprints attract more attention on social media, expediting scholarly communication. Numerous studies reveal positive correlations between mentions on X and paper citations \cite{eysenbach_can_2011, sudah_twitter_nodate, klar_using_2020}, underlining the influence of online platforms in academic impact.


In a randomized control trial (RCT) by \citet{luc_does_2021}, 112 cardiothoracic surgery articles from leading journals were randomized 1:1, resulting in an average of $3.1 \pm 2.4$ citations for tweeted articles after 1 year, compared to $0.7 \pm 2.4$ for non-tweeted ones. Conversely, \citet{tonia_if_2016} conducted a similar RCT for the International Journal of Public Health (IJPH) and found no differences in citations or article downloads between the exposure and control groups.

\citet{gunaratne_tweeting_2020} go beyond analyzing the impact of tweeting. They investigate whether authors tweeting about their papers in pulmonary and critical care medicine receive more citations than tweets from other accounts. The authors find that tweets from paper authors lead to 1.41 times more citations in one year and 1.51 times more in total. In a similar study on shoulder and elbow surgery publications, \citet{sudah_twitter_nodate} find no significant increase in citations when papers are tweeted by the authors.



The effect of tweeting remains unexplored in computer science and physics. 
We quantify the impact of tweeting on the number of citations and examine the impact of authors tweeting about their work versus tweets from other accounts.


\subsection{Novel measures of scientific impact (RQ5)}
We also study the effects of our treatment on more contemporary measures of paper-level impact such as highly influential citations \cite{ValenzuelaEscarcega2015IdentifyingMC}. Semantic Scholar employs a machine learning algorithm to calculate this metric, which assesses a citation's significance by analyzing contextual information extracted from the paper's content. The features used to classify a citation as highly influential include the text right before the citation, the section where the citation occurs, author overlap between the citing and the cited paper, etc. Therefore, the total number of highly influential citations may be a better measure of research impact. Thus, we hypothesize that they are relatively less affected by promotion on X and ArXiv than the plain number of citations.

\subsection{Differences in publishing trends between computer science and physics (RQ6)}

ArXiv's usage statistics \cite{ArxivUsageStats} reveals a significant increase in monthly submissions from January 2014 to March 2024: for physics, from approximately 1,432 to 2,566, and for computer science, from around 1,373 to 8,895, marking a 3.5-fold rise compared to physics. This prompts inquiry into the differing publication dynamics between these fields, including variations in the impact of social promotion. Prior studies have typically relied on RCTs which restricted their scope to individual publication venues or areas, whereas our approach using observational data to calculate effects, allows for the examination of publication dynamics and efects of promotion across computer science and physics.

\section{Data}

\begin{figure}[!t]
\centering
  \includegraphics[width=0.98\linewidth]{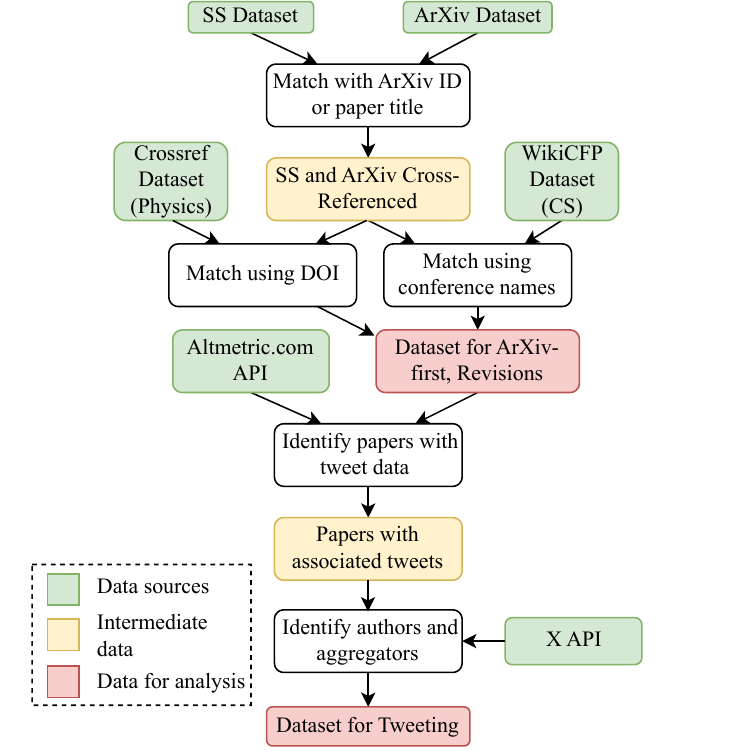}
  \caption{Step-by-step description of our dataset development process. ``SS'' stands for Semantic Scholar. The colored boxes stand for input/output data and the white boxes depict actions taken to produce the output.}
  \label{fig:data_flowchart}
\end{figure}
We create our dataset using data from six sources. In this section, we describe our process of dataset development (detailed in Fig. \ref{fig:data_flowchart}). Then, we describe the distribution of the number of citations of papers belonging to that dataset and how we use the dataset to answer our research questions.

      
                        
    

\subsection{Dataset development} 
We have created our dataset using data from six sources: Semantic Scholar\footnote{www.semanticscholar.org (CC BY-NC license)}, WikiCFP\footnote{www.wikicfp.com}, Crossref\footnote{www.crossref.org}, ArXiv\footnote{arxiv.org (CC0: Public Domain license)}, 
Altmetric.com, and the X API. Semantic Scholar is a search engine for scientific literature developed by the Allen Institute of AI. We used their publicly available dataset to get information about articles, authors, citations, and highly influential citations. WikiCFP is a website that contains details about calls for papers for scientific and technological conferences. We use the data from WikiCFP as a source of important dates relating to the submission and decision deadlines for conferences in computer science. Similarly, we collect the dates of journal publications for physics papers using the Crossref API. To access information about paper submission and revision dates, we utilize the dataset provided by ArXiv.
We examined citations spanning a five-year timeframe from the date of first publication (either on ArXiv or in a conference), incorporating papers up to 2017 in our analysis\footnote{Association for Computational Linguistic's (ACL) anonymity policy started in October 2017 so we have safely ignored that and have not taken the anonymity policy into account in our analysis.}, as we used Semantic Scholar data up to 2022. Altmetric.com provided us with research access to their proprietary dataset that serves as the data and metadata source for tweets related to each research paper in our dataset. Utilizing the unique tweet ID provided by Altmetric.com, we retrieve the tweet, its metadata, and the author's metadata using the X API.

\subsubsection{Cross-referencing Semantic Scholar and ArXiv.} The ArXiv metadata of computer science papers is cross-
referenced with the data from Semantic Scholar by matching the ArXiv IDs or the titles of the articles.

\subsubsection{Cross-referencing dates for conference and journal publications.} We use WikiCFP as a source of the following 5 dates for computer science conferences: the submission deadline, the notification date (the date when the author is notified about acceptance or rejection), final version (camera-ready) due date, and conference start and end dates. We noticed that a few top conferences and dates for some important conferences in computer science (like NeurIPS, AAAI, and ACL conferences) were not present on WikiCFP. We manually added these missing conferences to our dataset and annotated the missing dates. Details about the annotation process can be found in Appendix B. Despite this, we could not find all 5 required dates for around 150 important conferences, e.g., some of the older NeurIPS editions. Overall, we have 46.5k conference editions with full information about them. 

We matched conference names between the WikiCFP and Semantic Scholar datasets to add information about conference dates for the articles on Semantic Scholar by extending the conference name matching algorithm used by \citet{demetrescu_which_2022}, but unlike their approach, we do not rely on DBLP for conference names. Instead, we utilize conference names and acronyms from WikiCFP, and conference names from Semantic Scholar.  Appendix C contains the details of our algorithm. Our final dataset after cross-referencing Semantic Scholar, WikiCFP, and ArXiv contains 18,113 articles with author and citation information (including highly influential citations), conference dates, and ArXiv submission data. 

For journal articles in physics, we used Crossref as a source for publication dates and cross-referenced this data with papers on Semantic Scholar using the paper's DOI to create a dataset that contains 501,766 articles. 

These datasets are used for calculating trends for ArXiv-first submissions and revisions (RQ1) and to calculate the ArXiv-first and revision effects (RQ2,3). 

\begin{figure}[t]
\centering
\includegraphics[width=0.48\textwidth]{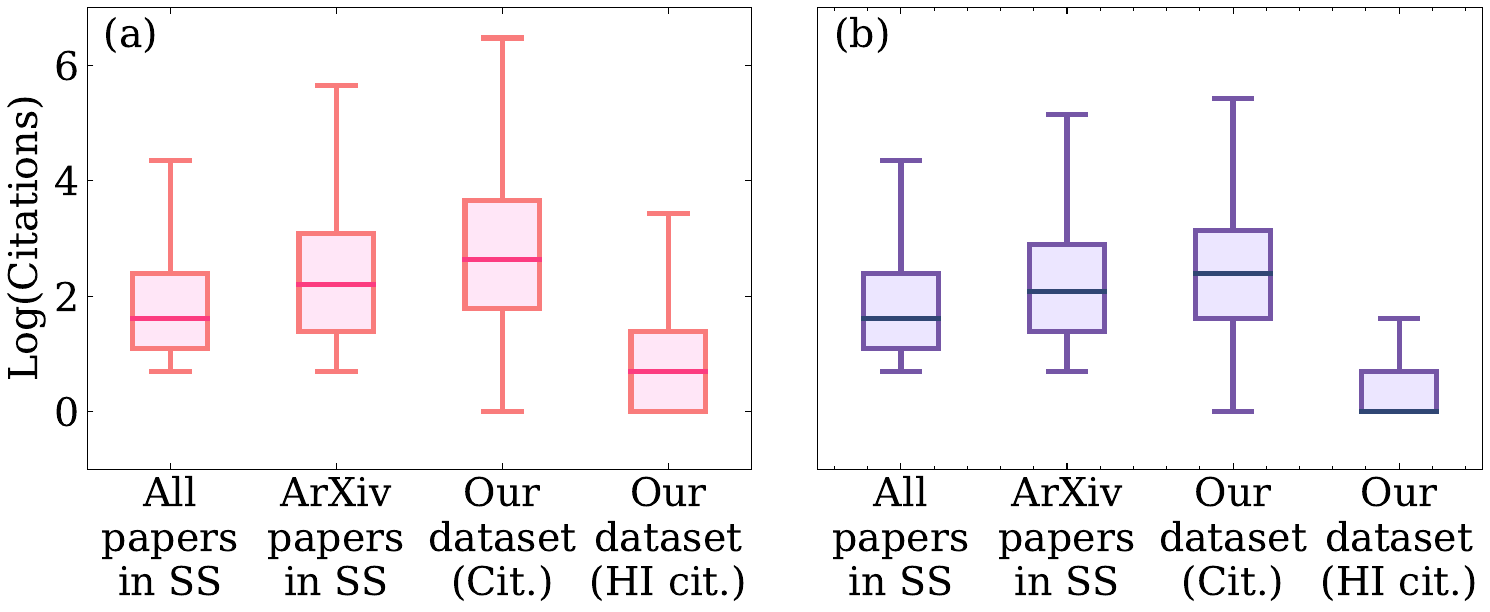}
\caption{Comparison of the distribution of citations in 5 years after the first date of publication for all papers on Semantic Scholar (SS), those on ArXiv cross-referenced with Semantic Scholar, and our dataset derived from cross-referencing ArXiv, Semantic Scholar, and WikiCFP/Crossref for (a) computer science and (b) physics. Cit. refers to citations and HI cit. refers to highly influential citations.} 
\label{fig:sampling_bias}
\end{figure}


\subsubsection{Identifying papers with tweets.} The Altmetric.com API provides rich information about papers that are publicized on social media, including fine-grained information about each social media post describing a paper. For computer science papers, using the API, we found 13,564 papers that have at least one tweet associated with them, 2,929 papers that have no tweets associated with them, and Altmetric.com did not contain information about 1,620 papers in our dataset. In physics, 134,563 papers have at least one associated tweet, while 188,692 have no tweets linked to them. Additionally, data for 178,511 papers could not be retrieved using Altmetric.com.

\subsubsection{Identifying papers tweeted by authors and aggregators.}

In this section, we discuss the algorithms used to identify papers tweeted by authors and aggregators. We define aggregators as X feeds curating research papers, including both bot accounts using keyword matching or other specialized algorithms on ArXiv and individuals curating papers in their X feeds. We aim to isolate the impact of tweeting (RQ4.1) itself from the distinction between whether the tweet originated from an author or an aggregator. Hence, when computing the treatment effect for tweeting, we factor in whether the paper was shared by an author or an aggregator. Also, to calculate the effect of author(s) tweeting (RQ4.2) we need information about the papers tweeted by authors. For this purpose, we must identify aggregator accounts and employ a name-matching algorithm to associate names between X users and paper authors.
\begin{itemize}
    \renewcommand\labelitemi{--}
    \item \textbf{Identifying aggregator accounts.} 
To identify aggregators, we set a threshold based on the frequency of tweeting research papers after a close manual inspection, classifying accounts exceeding one paper every 10 days as aggregators. We integrate data on these aggregators' tweets and followers as a control variable in our analysis.
 \item \textbf{Matching authors names to X profiles.} 
We identify papers that authors tweeted about by cross-referencing author names with X users who tweeted about a paper. We employ a name matching algorithm and estimate false negatives using an independent validation classifier. 
Our name matching algorithm identifies 1,344 papers tweeted by authors for computer science and 13,812 for physics, with a true positive rate of 94\%. We next try to estimate the prevalence of false negatives in our name matching algorithm. Employing a method from \citet{mane2005estimating}, we estimated approximately 170 false negatives for computer science, constituting about 1\% of papers in our dataset, which we conclude is negligible.
Appendix D contains details on the algorithm and false negative estimation.
\end{itemize}

This dataset is used to find trends in tweeted papers over the years  (RQ1) and to calculate the effects of tweeting (RQ4.1, 4.2).

\subsection{Dataset characteristics}
Publications in our dataset have a higher number of mean and median citations compared to all computer science or physics papers in Semantic Scholar or ArXiv, since only the higher-tier  conferences and journals appear in WikiCFP and Crossref (compare first through third boxes in Fig.~\ref{fig:sampling_bias}(a) or ~\ref{fig:sampling_bias}(b)). 
Our estimates of treatment effects are for papers that are published in such high-tier conferences and journals. We note that for lower-tier conferences and journals, treatment effects will be proportionally lower.
Finally, we notice that the number of influential citations is about 7 times smaller than the number of plain citations for computer science and 20 times smaller for physics (third and fourth box plots in Fig.~\ref{fig:sampling_bias}(a) and ~\ref{fig:sampling_bias}(b)). 

We use our entire dataset for computer science and physics to study temporal trends in paper promotion (RQ1). To calculate the effects (RQ2-5), we use all 18,113 papers in computer science and a 10\% subsample of the physics dataset which includes 50,177 articles. We opt for this compromise due to resource constraints, as we lacked the capacity to train superlearners containing tree-based models like BART on the entire physics dataset, since these models are not space-efficient and require more than 200 GB of RAM to run a single experiment.



\section{Methodology for treatment effect estimation}

\begin{table}[!t]
    \centering
    \resizebox{0.95\columnwidth}{!}{%
    \begin{tabular}{ll}
    \toprule
    \bf Control Variable & \bf Reason to Control  \\
    \toprule
    
     \textit{\textbf{ArXiv-first effect}} \\
    \text{Top author: Total number of citations}  & \\
     Top author: h-index & Self-selection bias\\
    First author: Total number of citations & (established author)\\
    First author: h-index& \\
    Number of authors  &  \\
    \midrule[0.5pt]
    Conference where the paper was   & Self-selection bias \\
    accepted & (quality)\\
    Conference rating (CS)/ &  \\
    Impact Factor of a journal (Physics) &  \\
    
    \midrule[0.5pt]
    Was the paper tweeted?  & Self-selection bias \\
    \midrule[0.5pt]
    Subject (CS, Physics etc.) & Area bias\\
    Computer Science subfield (AI, & \\
     Networking, etc.) & \\
    \midrule[0.5pt]
    Publication date & Temporal bias\\
    \arrayrulecolor{black}
    \midrule[1pt]
    
    \textit{\textbf{Revision effect}} \\
    +Paper submitted to ArXiv before  & Revision Effect  \\ 
    or after conference date & might be correlated  \\
    & with ArXiv-first effect\\
    \midrule[1pt]
    \textit{\textbf{Effect of tweeting}} \\
    +Sum of followers of X users who & Audience size\\
    tweeted about a paper & \\
    \midrule[0.5pt]
    +Tweeted by author & Exposure through\\
    +Tweeted by aggregator & authors or aggregators \\
    & on X \\
    \midrule[0.5pt] 
    +Was the paper revised on ArXiv? & Effect of tweeting  \\ 
        & might be correlated  \\
        & with Revision \\
    \midrule[0.5pt] 
    -Was the paper tweeted? & Removing the treatment \\
    & variable \\
    \midrule[1pt]
    \textit{\textbf{Effect of author(s) tweeting}} \\
    +Sum of followers of authors on X &  Audience size\\
    +Sum of followers of aggregators on X & \\
    \midrule[0.5pt]
    +Tweeted by scientist  & Exposure through\\ 
    +Tweeted by science communicator  & other channels \\
    +Tweeted by practitioners & on X\\
    +Tweeted by members of the public & \\
     \midrule[0.5pt] 
    -Sum of followers of X users who  & Audience size already  \\
    tweeted about a paper & captured using\\
    & the above control variables \\
    \midrule[0.5pt]
    -Tweeted by author &  Removing the treatment\\
    & variable \\
    -Tweeted by aggregator & \\
    \bottomrule
    \end{tabular}%
    }
    \caption{Control variables and the reason to control for the variable. The set of control variables used to estimate the ArXiv-first effect is also used when estimating the other effects (let us call this set $S$). We cumulatively add or remove new control variables with a "+" or "-" in front of their names, respectively. For instance: when estimating revision effect, we use 1 additional feature (over $S$); when estimating tweet effect, we add 5 additional variables and remove 1 (in comparison to $S$).}
    \label{tab:confounders}
\end{table}

This section describes the methods used for treatment effect estimation, utilized control variables and the rationale behind using them, followed by details about feature selection and the models used for treatment effect estimation.

One can estimate a treatment effect by modeling outcome variable, $Y$, and applying standardization, or by modeling treatment variable $A$, and applying inverse probability weighting, while controlling for all confounders $W$~\cite{hernan2020causal}. 
We refer to the regression model of outcome as $\mathbb{\hat{E}}[Y|a,w]$, where $W$ are control variables. The treatment model corresponds to so-called propensity scores $\hat{P}(A=1|w)$.
An estimate of treatment effect relies on the specification of respective model. If model specification is correct, then the estimate is unbiased. Doubly robust methods estimate both the model of treatment and outcome, so we have two shots at using the right model. If one of the models is correct and matches reality, then the estimate is unbiased. To further diminish the probability of model misspecification, for our treatment and outcome models we train a superlearner~\cite{van2007super}, a weighted ensemble of multiple machine learning sub-models. Superlearner is trained to provide  higher weights to the sub-models that generalize better.

The treatment variables, $A$, are the ones defined in RQ2-4. Our outcome variables here, $Y$, are the number of total citations (RQ2-4) and the number of highly influential ones (RQ5). We consider citations in the first five years since the initial publication (on ArXiv or in a peer-reviewed venue). By doing this, we can compare older papers, that might have higher citation counts as they were published earlier, directly with newer ones, that is we control for the early access. 

Finally, we control for multiple confounders, $W$. Table \ref{tab:confounders} details all control variables. First, we describe the core set of control variables, used for calculating all the effects (RQ2-4). Then, we describe the adjustment in control variables for calculating the effect of tweeting (RQ4). 


\subsection{Control variables}

The self-selection postulate~\cite{kurtz_effect_2005} states that better papers or papers from established authors are published more on ArXiv. This is referred to as the paper quality bias and the established author bias in this paper, respectively. To control for the established author bias, we use the total citation counts and h-index of the first author and the top author (selected based on maximum total citation counts). We use the top 150 conferences or journals (estimated based on the number of papers in each conference in our dataset) as a one-hot encoded feature. We also control for the average number of citations of papers published per conference, and the conference rating from the GII-GRIN-SCIE Conference Rating Initiative\footnote{https://scie.lcc.uma.es:8443/}. For physics venues, we control for the impact factor of the journal calculated using data from Scimago\footnote{https://www.scimagojr.com/}. These control variables give us a signal about the quality of the paper and hence can be used to control for the quality bias.

We note that tweets by publication authors can be treated as self-selection signals. Thus, we use the presence of a tweet about a paper as another control variable. 
We fetch total tweet counts from Altmetric.com for each paper and encode it into a binary variable indicating whether a paper was tweeted or not.~\citet{eysenbach_can_2011} finds that majority of tweets are made on the day of publication or the day after. Therefore, knowing that a paper was tweeted (i.e., if the tweet count is greater than zero) is a good indicator that a paper was tweeted right after publication and encompasses both axioms of the self-selection bias (i.e., papers may be tweeted by an established author and/or due to their quality).


We also use information about control variables like the subject category of the paper other than computer science or physics (like math, biology etc.) if it exists, the computer subfield it belongs to (like artificial intelligence and databases for computer science, and condensed matter and astrophysics for physics), and the venue it was published in to control for biases arising from certain areas being more popular at different periods of time and hence accruing more citations. We call this area bias. The publication date is also used as a control variable to adjust for temporal changes. 

\subsection{Differences in control variables across treatments}
 When estimating the tweeting effect (RQ4), we control for a few additional variables. We describe these variables and the need to control for them in this section. First, we control for the total number of followers of X users who tweeted about the paper. 
In this way, we control for the impact of differing X audience sizes.
When we calculate the effect of tweeting by authors, we also control for the sum of followers of the authors and aggregators to control for the same bias. Altmetric.com also provides demographic information about people tweeting about a paper dividing X users into scientists, science communicators, practitioners, and members of the public. We use four binary variables to indicate whether the paper was tweeted by each of these demographic groups and use this to control for exposure through other channels except authors on X.



\subsection{Feature selection}
We perform feature selection using Variance Inflation Factor (VIF) to remove multicollinearity among any of the features in the model. We remove the average citations for a conference as a feature as it is collinear with other features. 

\begin{figure}[t]
\centering
  \includegraphics[width=0.85\linewidth]{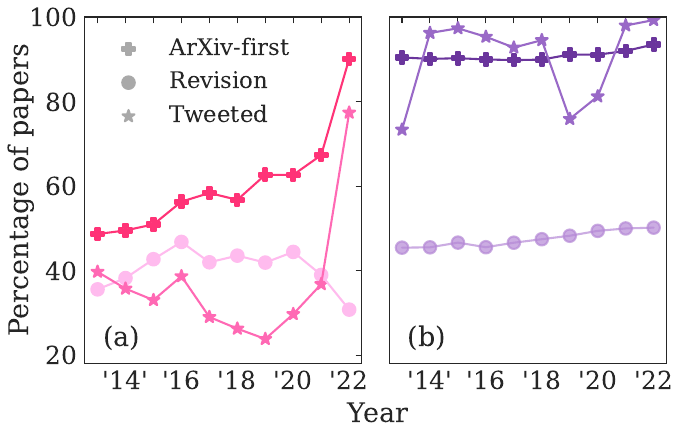}
  \caption{Percentage of papers that are submitted first and revised on ArXiv, and tweeted on X for (a) computer science and (b) physics.}
  \label{fig:percentage_papers}
\end{figure}

\begin{figure*}[t]
\centering
\includegraphics[width=0.9\textwidth]{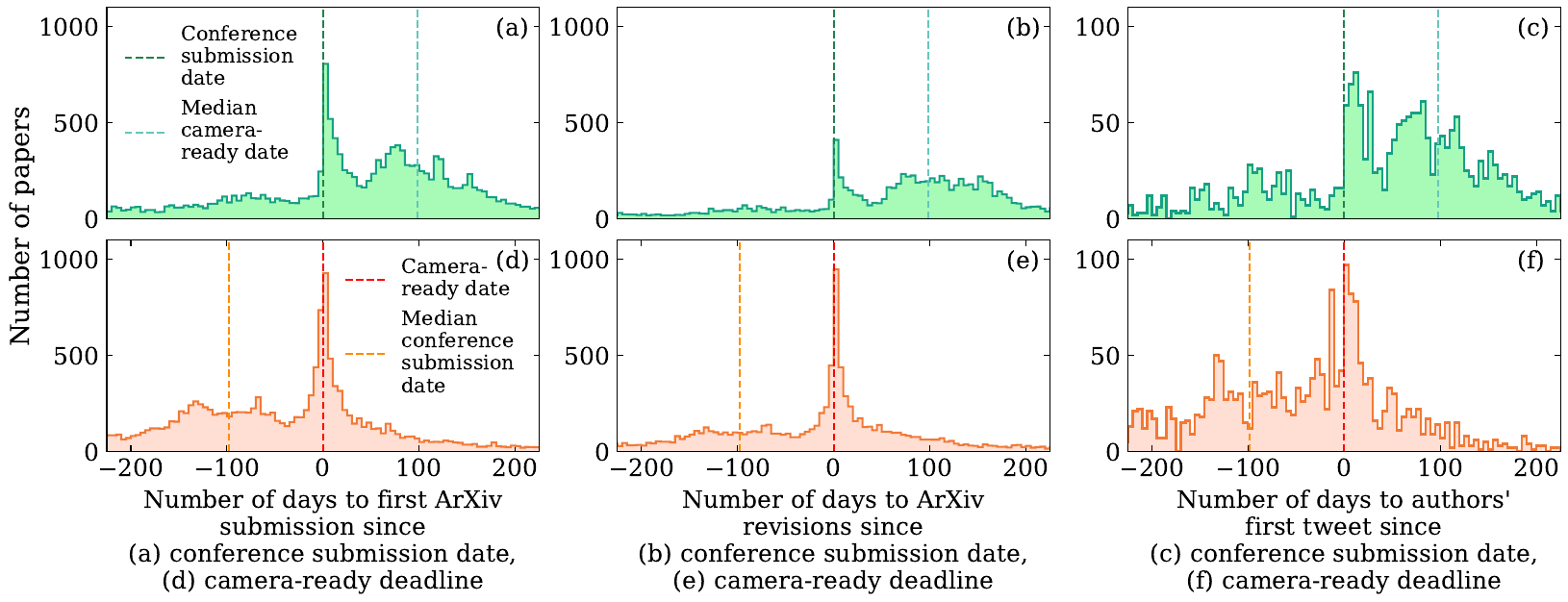}
\caption{When do ArXiv and X or Twitter publications happen for computer science? Histogram of the number of days since  (a) the conference submission date, (d) the camera-ready deadline to the first ArXiv submission, histogram of the number of days since (b) the conference submission date, (e) the camera-ready deadline to ArXiv revisions, and histogram of the number of days since the (c) the conference submission date, (f) the camera-ready deadline to the first tweet by an author.}
\label{fig:first_submissions_revisions_temporal}
\end{figure*}

\subsection{Models}

To increase the probability that our estimates of average treatment effects are unbiased, we use doubly-robust Targeted Maximum Likelihood Estimation (TMLE)~\cite{schuler2017targeted}. TMLE is semi-parametric and makes minimum assumptions about the distribution of the data.

The TMLE package in R allows the usage of superlearners. Superlearners are ensemble models that use $k$-fold cross-validation to weight $n$ base learners. The outputs of the cross-validation sets from $n$ base learners serve as independent variables that are passed on to a metalearner, typically a linear regression model, with true outputs of the cross-validation sets serving as the dependent variable. Following training, the $n$ coefficients of the metalearner determine the weights for each base learner.
Our outcome model $\mathbb{\hat{E}}[Y|a,w]$ is a superlearner that utilizes Bayesian Additive Regressive Trees (BART) and random forest regression. TMLE uses an additional treatment model making the approach doubly robust. The treatment variable is always binary and is estimated using a superlearner $\hat{P}(A=1|w)$ trained on a mixture of a logistic regression model, Discrete Bayesian Additive Regressive Trees (DBARTS) and Generalized Additive Models (GAM). After training the treatment and outcome models, TMLE performs an additional targeting step using both the models which optimizes the bias-variance tradeoff for the average treatment effect instead of the bias-variance tradeoff for the outcome which is common in other doubly robust estimators. This leads to a new adjusted expected outcome $\mathbb{\hat{E}}^*[Y|a,w]$, which can be used to estimate average treatment effect.
All models are initialized using default parameters in R. The superlearner models employ 10-fold cross-validation to find the best weights for each model.

\section{Results}
\label{sec:ate}




\subsection{Temporal trends in paper promotion (RQ1)} 


We are interested in understanding how the fractions of papers submitted early or revised on ArXiv
change over time.
The percentage of papers submitted before the conference to ArXiv has increased from 48.6\% in 2013 to 90.1\% in 2022 (Fig.~\ref{fig:percentage_papers}(a)) for computer science. In fact, submitting early to ArXiv has become the norm in computer science given the fast-moving nature of the field. The percentage of papers that are tweeted (by authors or any other account) also shows a similar trend with an exponential increase in the past few years for computer science, reaching 78\% in 2022. On the contrary, 90.6\% of physics papers in 2013 were being submitted to ArXiv before the journal publication and this percentage went up to 93.5\% in 2022 (Fig.~\ref{fig:percentage_papers}(b)). The percentage of papers tweeted for physics lies in the 80\% to 99\% range and does not grow fast like for computer science.
The percentage of computer science papers revised on ArXiv 
remains relatively constant with minor fluctuations throughout the years, whereas in physics increases by about 6\% between 2013 and 2022.

As we have fine-grained information about dates of computer science conference submissions and camera-ready deadlines, we also analyze \textit{when} researchers make their first submissions to ArXiv and \textit{when} they submit a revised manuscript on ArXiv. Fig.~\ref{fig:first_submissions_revisions_temporal} shows that there are two major peaks corresponding to the periods when researchers make their first submission to ArXiv: one at the conference submission date and the other at the final version due date (camera-ready deadline) (Fig. \ref{fig:first_submissions_revisions_temporal}(a), \ref{fig:first_submissions_revisions_temporal}(d)). We see a similar trend for revisions, with researchers submitting revised versions of their transcripts much more often around the time of the final version due date than the conference submission deadline (Fig. \ref{fig:first_submissions_revisions_temporal}(b), \ref{fig:first_submissions_revisions_temporal}(e)), 
likely after incorporating changes that the reviewers point out in the peer-review process. 
We also see similar trends for the first  tweet by an author(Fig. \ref{fig:first_submissions_revisions_temporal}(c), Fig. \ref{fig:first_submissions_revisions_temporal}(f)) where there is a peak at the conference submission date and the final version due date but this occurs only because 99.02\% of papers tweeted by authors in our dataset receive their initial tweet from an author on the same day as their initial ArXiv submission. 

\begin{table*}[!t]
    \centering
    \resizebox{0.99\textwidth}{!}{%
    \begin{tabular}{c|c|ccccc|ccccc}
    \toprule
     \multirow{2}{*}{\textbf{Outcome}} & \multirow{2}{*}{\textbf{Treatment}} & \multicolumn{5}{c|}{\pgfsetfillopacity{0.8}\colorbox{Pink1}{\textbf{Computer Science}}} & \multicolumn{5}{c}{\pgfsetfillopacity{1}\colorbox{Plum3}{\textbf{\textbf{Physics}}}} \\ 
   \cmidrule{3-12}
    & & \bf  ATE & \bf  Relative & \bf p-value & \bf Cross-val & \bf AUC & \bf  ATE &   \bf  Relative & \bf p-value & \bf Cross-val  &  \bf AUC \\ 
   &  & \bf  & \bf ATE &  & \bf $R^2$ &   & \bf  & \bf ATE & & \bf $R^2$ &  \\ 
    \midrule
    
 \multirow{6.5}{*}{{\textbf{All citations}}} &  \text{ArXiv-first effect}  & $21.1 \pm 17.4$ & 0.30 & 0.02 & 0.14 &	0.77 &  $3.9 \pm	1.1$	 & 0.20 & $<10^{-6}$ & 	0.13	& 0.75  \\
    \cmidrule{2-12}
    & \text{Revision Effect} & $18.4	\pm 17.6$ & 0.26 & 0.04 &  0.13	& 0.73 &  $4.3	\pm 0.9$	& 0.22 & $<10^{-6}$	& 0.13	& 0.72  \\
    \cmidrule{2-12}
    & \text{Effect of Tweeting} & $44.4	\pm 8.0$ & 0.58 & $< 10 ^{-6}$ & 0.19 &	0.97 &  $6.9 \pm	3.5$ & 0.29 &	$8 \times 10^{-5}$ &	0.23 & 0.99 \\
    \cmidrule{2-12}
    & \text{Effect of Author(s) } & $28.4	\pm 11.2$ & 0.32 & $< 10 ^{-6}$	& 0.18	 & 0.99 &  $0.8 \pm 3.9$ &	0.03 & 0.74	& 0.24 & 0.99  \\
    &\text{Tweeting} & & & & & & & \\
    
    \cmidrule{1-12}

     \multirow{5}{*}{{\textbf{Highly influential}}} &  \text{ArXiv-first effect}  & $3.4 \pm 2.9$ & 0.36 & 0.02 & 0.07 &	0.76 &  $0.2 \pm 0.1$ & 0.21 & $2 \times 10^{-5}$ & 0.14 & 0.75  \\
    \cmidrule{2-12}
    \multirow{4}{*}{{\textbf{citations}}} & \text{Revision Effect} & $2.5	\pm 4.0$ & 0.27 & 0.21 & 0.07 	&  0.72  & $0.2	\pm 0.1$	& 0.21 & $<10^{-6}$ & 0.15 & 0.72 \\
    \cmidrule{2-12}
    & \text{Effect of Tweeting} & $7.6	\pm 1.61$ & 0.68 & $< 10 ^{-6}$ &  0.10 & 0.98 & $0.4 \pm 0.2$ & 0.33 & $2 \times 10^{-4}$ &	0.18 &	0.99  \\
    \cmidrule{2-12}
    & \text{Effect of Author(s) } & $6.2	\pm 2.5$ & 0.48 & $< 10 ^{-6}$	& 0.11 & 0.99 & $0.3 \pm 0.1$ & 0.25 & $<10^{-6}$ & 0.17 & 0.99 \\
    & \text{Tweeting} & & & & & & & \\
    [1ex]
    \bottomrule
    \end{tabular}}
    \captionof{table}{Average treatment effect (ATE estimated via TMLE) of each of our treatments on all citations and highly influential citations over 5 years after first publication. ``Relative ATE'' is the ATE divided by the mean number of citations or highly influential citations for that discipline.}
  \label{tab:ate_results_5_years}
\end{table*}

\subsection{Average treatment effects (RQ2-RQ4)}

We find statistically significant positive effects for the four treatments for computer science (CS) using TMLE, suggesting that submitting early, revising, tweeting, and the author(s) tweeting lead to a higher number of citations over 5 years after publication. For physics, we also find positive effects, although effect sizes are smaller.
The results are summarized in Table~\ref{tab:ate_results_5_years}.

For the ArXiv-first effect, we notice a positive average treatment effect ($21.1 \pm 17.4$ for CS, and $3.9 \pm 1.1$ for physics), which suggests that submitting to ArXiv before conference deadline on average adds $21$ citations for CS and $4$ citations for physics in the first 5 years from the initial publication.
It is possible that ArXiv, with its millions of subscribers, provides a visibility exceeding that of conference proceedings.
The positive ArXiv-first effect could also indicate a significant first-mover advantage \cite{newman2009first} where a researcher putting their work out early on ArXiv would get higher number of citations compared to others who wait for the completion of the conference peer-review process, which can take several months, disrupting paper's timeliness and hotness in a fast-paced research area.

The effect of revisions is also positive ($18.4 \pm 17.6$ for CS, and $4.3 \pm 0.9$ for physics). This result suggests that incorporating feedback from research community leads to greater impact or that important research papers are more likely to be revised with time, even post-publication, leading to more citations.


We see a positive treatment effect for tweeting as well ($44.4 \pm 8$ for CS, and $6.9 \pm 3.5$ for physics) which suggests that tweets are an impactful tool for publicizing one's work in the age of the internet. Research paper showing up on X tend to garner much more citations than papers that are not tweeted.

Among the papers that are tweeted for CS, there is also a positive treatment effect if it is tweeted by an author ($28.4 \pm 11.2$ for CS), as compared to some other account.  This suggests that being active on social media, sharing, and promoting one's work leads to greater visibility and, ultimately, more citations. This effect is not statistically significant using TMLE for papers in physics.

We observe the same qualitative results for average treatment effects when we use standardization instead of TMLE. One notable exception is that we get a statistically significant positive effect for authors' tweets for physics ($3.1 \pm 1.0$) with this method. The results for standardization can be found in Appendix E.


\begin{figure}[!htb]
\centering
\includegraphics[width=0.47\textwidth]{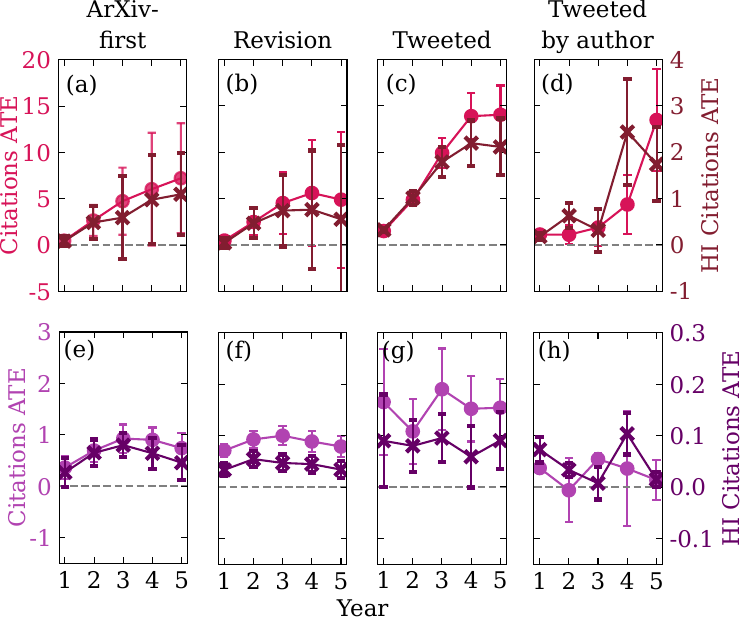}
\caption{TMLE Average Treatment Effects (ATE) computed separately for each year over the first 5 years after publication for (circles) standard citations and (crosses) highly influential citations to (a-d) computer science and (e-h) physics papers.
Error bars represent 95\% confidence intervals. Detailed results in Appendix F.
}
\label{fig:ate_over_years}
\end{figure}

\subsection{Treatment effects for influential citations (RQ5)} 
Highly influential citations have positive but smaller effect sizes, given their lower numbers. The ratio of citations to highly influential citations is approximately 7:1 for computer science and 20:1 for physics.
To account for these differences in the number of citations between our two disciplines and types of citations, we compute \textit{relative ATE} as ATE divided by the number of citations of the respective type for that discipline. 
Computer science papers average 69.57 citations and 9.37 highly influential citations, while physics papers average 19.20 citations and 0.97 highly influential citations.\footnote{The disparities between disciplines appear because WikiCFP provides all required dates only for top computer science venues, whereas our physics dataset contains many more papers from less prestigious venues, such as Physical Review A, B, C, D, and E.}
Relative treatment effects helps us account for these differences when making comparisons across citation types and disciplines.
For instance, the relative ATE is nearly the same for standard and highly influential citations within a discipline (compare relative ATE values between the top and the bottom of Table~\ref{tab:ate_results_5_years}), likely because relatively speaking the effects are the same for standard and influential citations.
The treatment effects for highly influential citations exhibit also nearly the same temporal trends as raw citation counts (Fig.~\ref{fig:ate_over_years}).
Overall, our hypothesis that this novel measure of scientific impact is less affected by tweeting does not hold.

\begin{figure}[!t]
\centering
    \includegraphics[width=0.99\linewidth]{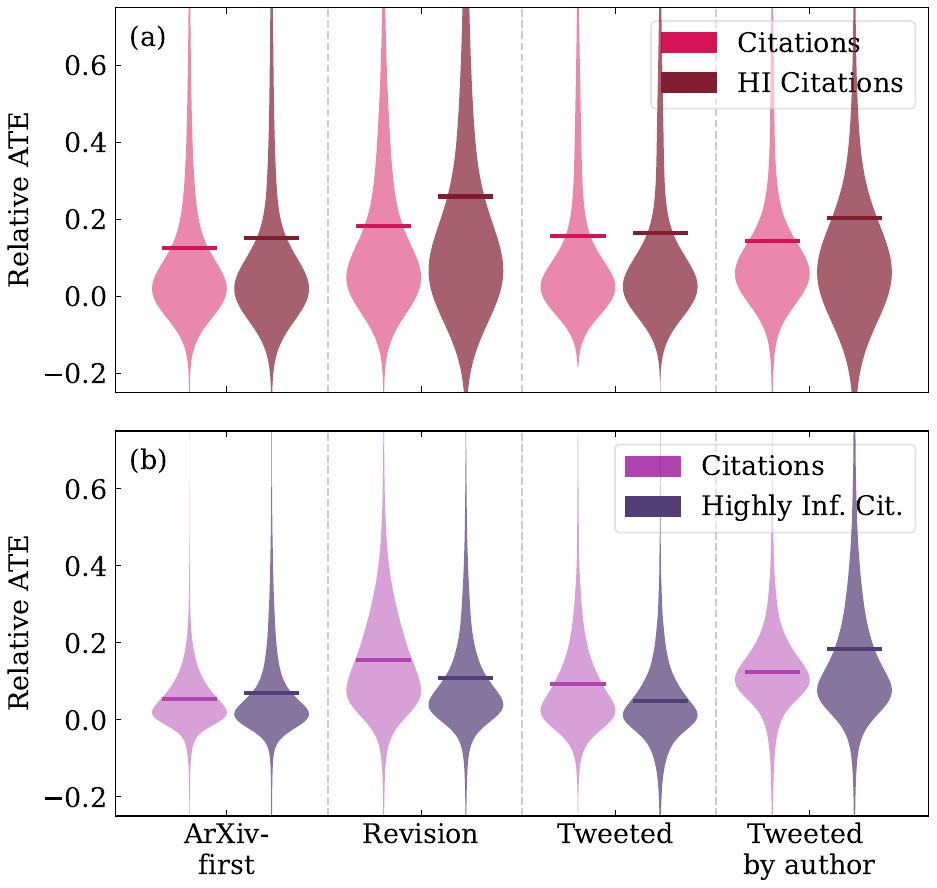}
    \captionof{figure}{Treatment effects for individual papers using standardization for (a) computer science and (b) physics. The ATE is marked with a solid line.}
  \label{fig:heterogeneous_effects}
\end{figure}

\subsection{Differences in effects between computer science and physics (RQ6)} 

\subsubsection{Average treatment effects.} 
We look at the relative ATE values to compare treatment effects in computer science and physics (compare left and right sides of Table \ref{tab:ate_results_5_years}). Notably, the differences in relative ATE values are most pronounced for treatments that are affected by social factors such as word-of-mouth spreading: the effect of tweeting (0.58 for computer science vs. 0.29 for physics, a 50\% decrease for physics) and the the ArXiv-first effect (0.30 for computer science vs. 0.20 for physics, a 33.3\% decrease). On the other hand, the revision effect, which is the least influenced by social factors, shows the smallest decrease for physics (0.26 for computer science vs. 0.22 for physics, only a 15.3\% decrease). These results suggest that citations in computer science are more affected by social influence than in physics, in line with the following findings.

\subsubsection{Distributions of treatment effects per paper.} To compute per-paper effects, we could not use doubly robust estimation, since it is designed for average treatment effect estimation. The long-tailed distribution of per-paper treatment effects highlights the diverse nature of these effects, indicating relative effect sizes vary among papers in our dataset based on individual characteristics (Fig.~\ref{fig:heterogeneous_effects}). Interestingly, computer science treatment effects display a wider spread than for physics. In other words, the effects of treatment on citation numbers is less predictable for computer science publications, possibly due to presence of a stronger social influence mechanism. This result aligns with the seminal experiments of Salganik at el. indicating that social influence in a (digital) market increases unpredictability of product success in that market~\cite{salganik2006experimental}.

\subsubsection{Average treatment effects over time.} 
Finally, we study how treatment effects change over years (Figure \ref{fig:ate_over_years}).
For computer science, the ArXiv-first and tweeting effects increase in strength over the first four years. Conversely, in physics papers, these effects do not seem to grow, particularly the effect of tweeting. Here, we also notice the same pattern as seen with the relative ATE values, i.e., the most notable variation occurs in treatments influenced by social promotion, i.e., the effect of tweeting (Figure~\ref{fig:ate_over_years}(c, g)), the effect of authors tweeting (Figure~\ref{fig:ate_over_years}(d, h)), and the ArXiv-first effect (Figure~\ref{fig:ate_over_years}(a, e)), compared to the revision effect that is least related to social promotion (Figure~\ref{fig:ate_over_years}(b, f)). These findings, along with the prior two results, suggest social influence cascades \cite{shakarian2015independent, pastor2015epidemic} boost citation rates in computer science papers, while being less influential in physics.

\section{Discussion}

 
We find positive effects for ArXiv-first submissions, possibly indicating a first-mover advantage \cite{newman2009first}.
This presents a challenge, as authors may publish preliminary results on ArXiv to claim ownership, accumulating more citations than more comprehensive studies on the same topic. 

ArXiv paper revisions also show positive effects, signaling efforts to address weaknesses or that the topic of study is important enough for the authors to keep working on the paper and improve it with time. These enhancements serve as strong indicators of improved quality, resulting in increased citations. 

We find a positive effect of tweeting on the number of citations, i.e., one's research gets more attention if it is promoted on X, highlighting social effects in dissemination of scientific information. This finding suggests that conferences prohibiting ArXiv submissions diminish the paper's potential impact on the community. 

Although the affordances outlined in this paper enhance the speed of scientific communication and are gaining popularity, they present challenges to the scientific review process. 
Our results suggest that 
millions of users who are on the lookout for new papers on ArXiv may lead to more exposures than peer-reviewed venues. However, unreviewed low-quality preprints on ArXiv can spread misinformation. 
One potential answer to this challenge is to adopt post-publication peer review \cite{ford2013defining, o2021overview}.
This approach not only facilitates the rapid dissemination of scientific information but also fosters meaningful discussions among researchers, thereby encouraging collaboration. If the research community adopts a system that allows organic and quicker ways for providing feedback, this could lead to new paradigms in scientific publishing, such as the ones enabled by systems such as OpenReview.net, where reviews can be posted instantly, visible publicly, and responded by authors. We anticipate such systems would quicken feedback loops, while at the same time present challenges for science evaluation.

We also notice significant differences in treatment effects between computer science and physics. These disparities indicate significant differences in the impact of social influence on citations within each field. Notably, citations in computer science exhibit a greater susceptibility to internet-based promotion, such as early submissions to ArXiv and tweeting, compared to physics. Conversely, treatments like revisions on ArXiv, which are not related to word-of-mouth spreading, yield similar effects in both disciplines. Additionally, citations in computer science appear to benefit from social cascading effects, evidenced by increasing treatment effects over yaers, in contrast to physics.
The pronounced social influence in computer science would also render the effects of our treatments on citation numbers more unpredictable, as illustrated by the wider distribution of per-paper treatments in Figure \ref{fig:heterogeneous_effects}. Overall, these three findings indicate an increased propensity for social conformity and collective behavior in computer science compared to physics, which may be related to the hotness, hype, and commercialization of research areas such as artificial intelligence. This hypothesis can be explored by future studies.

Citations and their more modern variants, that is highly influential citations, are volatile and can be easily influenced by promotion on ArXiv and X. Therefore, there is a need for measures of research impact that are impervious to these effects. 
Proposed citation count normalization methods, like field-normalized or network-normalized counts \cite{waltman2016review, ke2023network, hutchins2016relative}, offer alternative metrics that are independent of publication year or field. In contrast, altmetrics, such as the ones reported by Almetric.com or PlumX, directly depend on a paper's social media attention. Our comprehensive analysis reveals the necessity for developing research impact metrics that incorporate the social aspect of sharing scientific information. This concept is also put forth by \citet{ke2023network}, advocating the exploration of network-normalized metrics on a modified citation graph that accounts for social factors in the dissemination of scientific information. Classifying citations as topical (based on topical similarities between the papers) or non-topical (based on social or professional connections between the authors) in the citation graph drawing from research on social ties \cite{10.1145/2615569.2615672,grabowicz2016distinguishing, grabowicz2013distinguishing}, offers a promising avenue since topical citations are expected to carry more relevance than non-topical ones when analyzing the research impact of a paper. Employing such a socially-informed measure would further enhance the accurate assessment of scholarly influence. 

Nevertheless, we believe any new metric for assessing scientific impact will inevitably fall prey to Goodhart's law, which states that, "When a measure becomes a target, it ceases to be a good measure." In other words, we expect that measures of scientific impact will keep evolving, and be gamed. We hope that this evolution will be driven by a scientific process motivated by studies like the one we developed.

%

\paragraph{Limitations.} 
In this paper, we have made a conscientious effort to address numerous control variables to minimize biases but there still might exist unaccounted control variables which may impact our results. Although we use doubly robust estimators, model specification issues could still introduce bias. 

\paragraph{Acknowledgements.} We thank Altmetric.com for granting us access to their dataset and Semantic Scholar for making their dataset publicly available.

\bibliography{citations}

\subsection{Ethics Statement}

We do not endorse researchers to blindly optimize for the number of citations rather than contributing to human knowledge. Furthermore, if scientists begin submitting unfinished low-quality papers to arXiv, we expect the audience of ArXiv to decrease in size and the treatment effects to diminish in value. Thus, our results depend on the decision-making of scientists and the threshold on manuscript quality that they use to determine a good moment for submitting an early preprint. Furthermore, we believe that it is more important that researchers consider broad real-world and ethical impact of their work, rather than the ephemeral value of the number of citations.

\renewcommand{\thesubsection}{\thesection.\arabic{subsection}}
\section*{APPENDIX}

\subsection{A. Distribution of all citations for each treatment}

\begin{figure}[h]
\centering
\subfloat{\includegraphics[width=0.126\textwidth]{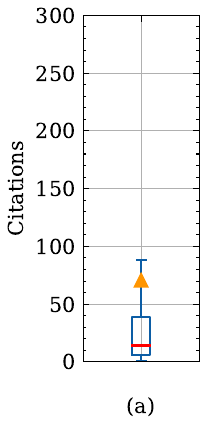}
\label{fig:subfig0}}
\subfloat{\includegraphics[width=0.08\textwidth]{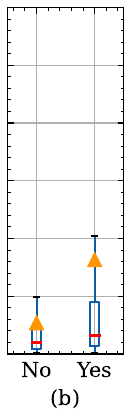}
\label{fig:subfig1}}
\subfloat{
\includegraphics[width=0.08\textwidth]{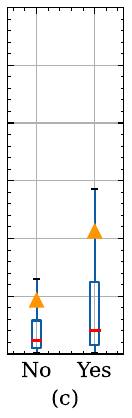}
\label{fig:subfig2}}
\subfloat{
\includegraphics[width=0.08\textwidth]{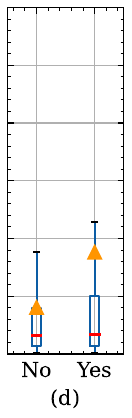}
\label{fig:subfig3}}
\subfloat{
\includegraphics[width=0.08\textwidth]{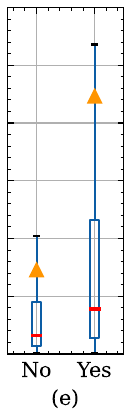}
\label{fig:subfig4}}
\caption{Box plots for the number of citations for each treatment for computer science. The red line in the middle of the box represents the median and the orange triangle represents the mean for (a) Number of citations for all papers, (b) ArXiv-first, (c) Revision, (d) Tweeted, and (e) Tweeted by author.}
\label{fig:box_plots}
\end{figure}

\begin{figure}[h]
\includegraphics[width=0.48\textwidth]{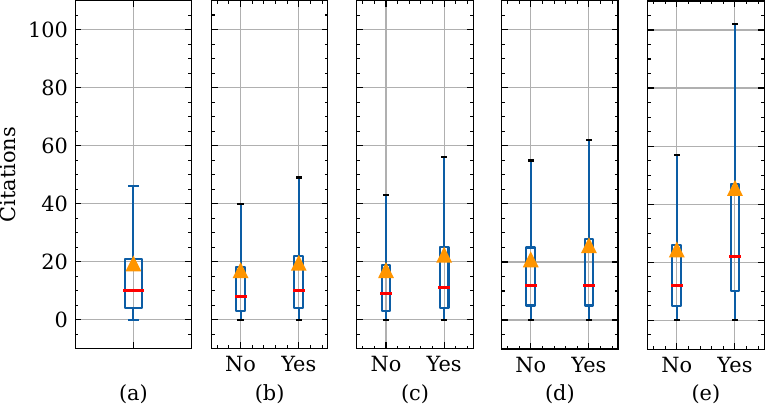}
\caption{Box plots for the number of citations for each treatment for physics. The red line in the middle of the box represents the median and the orange triangle represents the mean for (a) Number of citations for all papers, (b) ArXiv-first, (c) Revision, (d) Tweeted, and (e) Tweeted by author.}
\label{fig:box_plots2}
\end{figure}

\subsection{B. Missing top conferences and dates from WikiCFP}

We gathered all conferences from CSRankings.org and found that 8 of the top conferences were missing from WikiCFP which we added to our dataset(listed below). In addition to these 8 conferences, we identified around 400 conference editions with missing or incomplete WikiCFP information. We annotated their dates manually, using information publicly available on the internet, to include as many important computer science venues as possible and to increase our dataset size. 

List of top conferences in computer science that are missing from WikiCFP but are listed on CSRankings.org:
\begin{enumerate}
    \item IMWUT: Proceedings of the ACM on Interactive, Mobile, Wearable and Ubiquitous Technologies
    \item RTAS: IEEE Real-Time and Embedded Technology and Applications Symposium
    \item VIS: IEEE Visualization
    \item S\&P: IEEE Symposium on Security and Privacy
    \item Eurosys
    \item USENIX ATC: USENIX Annual Technical Conference
    \item SIGSOFT FSE: SIGSOFT Foundations of Software Engineering
    \item USENIX Security
\end{enumerate}

\subsection{C. Algorithm to match conference names between WikiCFP and Semantic Scholar}

\begin{algorithm}[h]
\caption{Text normalization}\label{alg:text_norm}
\begin{algorithmic}
\State \textbf{Input:} String $s$
\State \textbf{Output:} Normalized string $s\prime$

\\

\State SubstringList $\leftarrow$ ['part of', 'held with', 'conjunction with', 'colocated with', 'collocated with']

\\

\State $s\prime \leftarrow$ Convert $s$ to uppercase
\State $s\prime \leftarrow$ Remove diacritical marks from $s\prime$
\State $s\prime \leftarrow$ Remove special characters from $s\prime$
\State $s\prime \leftarrow$ Remove stopwords from $s\prime$
\State $s\prime \leftarrow$ Remove double spaces from $s\prime$

\\

\If{any substring $sub$ from SubstringList present in $s\prime$}
    \State $s\prime \leftarrow$ Drop trailing tokens in $s\prime$ starting from $sub$
\EndIf

\State return $s\prime$
\end{algorithmic}
\end{algorithm}

\begin{algorithm}[h]
\caption{Loose string match}\label{alg:loose_match}
\begin{algorithmic}
\State \textbf{Input:} String $s1$, String $s2$
\State \textbf{Output:} Percentage match $p$
\\
\State StringMatchCount $\leftarrow$ 0
\State s1Tokens $\leftarrow$ Get every word in $s1$ by splitting at every \indent \indent \indent  blankspace
\State s2Tokens $\leftarrow$ Get every word in $s2$ by splitting at every \indent \indent \indent blankspace
\\
\For{every $token$ in $s1Tokens$}
    \If{$token$ in $s2Tokens$}
            \State remove $token$ from $s2Tokens$
            \State StringMatchCount += 1
    \EndIf
\EndFor
\State p $\leftarrow$ StringMatchCount/len(s1Tokens)
\State return p
\end{algorithmic}
\end{algorithm}

\begin{algorithm*}[h]
\caption{Conference name matching}\label{alg:conf_name_match}
\begin{algorithmic}
\State \textbf{Input:} Semantic Scholar venue name $ssNames$, WikiCFP venue acronyms $wikiAcronyms$, dictionary containing WikiCFP  acronyms to venue names  $wikiAcronymsToNames$
\State \textbf{Output:} List of matches $match$
\\

match $\leftarrow$ []

\For{every $ssName$ in $ssNames$}
    \For{every $wikiAcronym$ in $wikiAcronyms$}
        \If {$ssName$ contains $wikiAcronym$}
            \If  {len($ssName$) $>$ 2}
                \State normalizedSSName $\leftarrow$ textNormalization  (ssName)
                \State normalizedWikiName $\leftarrow$ textNormalization  ($wikiAcronymsToNames[wikiAcronym]$)
                \State stringMatchPercentage $\leftarrow$ looseStringMatch  (normalizedSSName, normalizedWikiName)
                \If {stringMatchPercentage $\ge$ 0.75}
                    \State match.append($ssName$, $wikiAcronym$)  
                \EndIf
                \Else 
                \State match.append($ssName$, $wikiAcronym$)
            \EndIf
        \EndIf
    \EndFor
\EndFor

\State return match
\end{algorithmic}
\end{algorithm*}

To match conference names between between Semantic Scholar and WikiCFP, we first create a dictionary that maps conference acronyms to conference names using data from WikiCFP. Next, we check if there is match between the WikiCFP conference acronym and the Semantic Scholar conference name (typically, Semantic Scholar conference titles incorporate the conference acronym within their names.). If there is a match and the Semantic Scholar conference name has less than 2 words in it (i.e., it contains the conference name and the year, for e.g., ACL 2014), we add the conference name pair to $match$,  our final list of matched conference names. If the Semantic Scholar conference name has more than 2 words in it, we perform a loose string match on the normalized strings for the Semantic Scholar and WikiCFP conference names. The algorithms for text normalization and loose string matching are defined in Algorithms 1 and 2 respectively. If the percentage match is greater than 0.75, we add the conference name pair to $match$.

\subsection{D. Name matching algorithm to find tweets from authors}

\paragraph{Name matching algorithm:} 
We obtain tweet IDs for each paper in our dataset from Altmetric.com, then retrieve tweets and metadata using the X API. We perform fuzzy name matching by first converting Unicode characters to ASCII and standardizing to lowercase, and excluding names shorter than 2 characters (to reduce the number of false positives). Utilizing the fuzzywuzzy Python library, we apply a 95\% threshold for matching authors' last names with X usernames (e.g., John Doe). If no match is found, we repeat the process with screen names (e.g., @johndoe). This yields 1,344 papers with matching author names or X profiles. From this set, a random sample of 50 papers contains 94\% true positives for the name matching algorithm.


\paragraph{Calculating false negatives for the name matching algorithm:} To find the number of false negatives with this method, we use the methodology outlined in~\citet{mane2005estimating} where two independent classifiers are used to estimate the number of false negatives for each classifier. The number of false negatives ($\hat{n}_{00}$), i.e., the number of true positives misclassified by both algorithms, can be estimated as

\begin{equation}
\hat{n}_{00} = \frac{n_{01} \times n_{10}}{n_{11}},
\label{eq:false_negatives}
\end{equation}

\noindent where $n_{10}$ is the number of true positives that are identified correctly by classifier 1 but misclassified by classifier 2, $n_{01}$ is the number of true positives that are correctly classified using classifier 2 but misclassified by classifier 1, and $n_{11}$ is the number of true positives correctly classified by both the classifiers.  

To perform this analysis, the second classifier (referred to as the validation algorithm) finds the tweets that start with the word "our" in our dataset. This follows the intuition that tweets
from authors discussing their papers often begin with this word. We also tried other algorithms such as finding tweets that contain the word "our" instead of strictly starting with it, and tweets that check for other strings like "we" and "our paper" but the percentage of true positives using these methods were low. We find that the number of papers for which both the name matching algorithm and the validation algorithm give us a positive result is 213. On manual inspection of 50 of these papers, we find that the intersection has 94\% true positives.  Therefore, the number of true positives at the intersection of the two algorithms is $213*0.94=200.22 (n_{11})$. The validation algorithm has 34 more positives that are not part of the positives for the name matching algorithm. Out of these,  26 are true positives ($n_{01}$). And we already know that the number of positives for the name matching algorithm is 1344 with 94\% true positives which makes the number of true positives for the algorithm to be $1344*0.94=1263.36$ ($n_{01}$). We use Eq.\ref{eq:false_negatives} to calculate the total number of false negatives and $\hat{n}_{00}$ in this case is 154.21 ($\approx$ 154). 

Using this analysis, we find that we have a total of 1344 tweets by authors out of which 94\% are true positives and we are missing around 154 other true positives that are misclassified by both the algorithms and 26 true positives that are correctly classified by the validation algorithm but misclassified by the name matching algorithm, summing up to a total of 170 estimated false negatives for the name matching algorithm. This number constitutes about 1\% of all papers for which we did not find an X profile match, so we conclude that it is negligible.

\begin{table*}[!htb]
    \centering
    \resizebox{2\columnwidth}{!}{%
    \begin{tabular}{c|c|ccccc|ccccc}
    \toprule
     \multirow{2}{*}{\textbf{Outcome}} & \multirow{2}{*}{\textbf{Treatment}} & \multicolumn{5}{c}\textbf{Computer Science} & \multicolumn{5}{c}{\textbf{Physics}} \\ 
   \cmidrule{3-12}
    & & \bf  Standardization &   \bf  Relative & \bf p-value & \bf Train  &  \bf Cross-val & \bf   Standardization &   \bf  Relative & \bf p-value & \bf Train  &  \bf Cross-val \\ 
   &  & \bf ATE & \bf ATE &  & \bf $R^2$ & \bf $R^2$  & \bf ATE & \bf ATE & & \bf $R^2$ & \bf $R^2$ \\ 
    \midrule
    
 \multirow{6.5}{*}{{\textbf{All citations}}} &  \text{ArXiv-first effect}  & $9.0 \pm 3.7$ & 0.13 &    $< 10 ^{-6}$ & 0.49 & 0.19 &  $0.9 \pm 0.3$ &  0.05 &$< 10 ^{-6}$	& 0.51 & 0.17   \\
    \cmidrule{2-12}
    & \text{Revision Effect} & $12.0 \pm 3.6$  & 0.17 & $< 10 ^{-6}$ & 0.48 & 0.14 &  $3.0 \pm 0.4$ & 0.16 &	$< 10 ^{-6}$ &	0.52 &	0.17  \\
    \cmidrule{2-12}
    & \text{Effect of Tweeting} &   $11.9 \pm 1.9$ & 0.16 &  $< 10 ^{-6}$ & 0.62 & 0.26 &  $2.0 \pm 0.2$ &	0.08 & $< 10 ^{-6}$	& 0.58 & 0.24 \\
    \cmidrule{2-12}
    & \text{Effect of Author(s) } &  $14.8 \pm 8.9$ & 0.17 & 0.001 & 0.61 & 0.24  &   $3.1 \pm 1.0$  & 0.12 & $< 10 ^{-6}$	& 0.63 & 0.24  \\
    &\text{Tweeting} & & & & & & & \\
    
    \cmidrule{1-12}

     \multirow{5}{*}{{\textbf{Highly influential}}} &  \text{ArXiv-first effect}  & $ 1.4 \pm 0.8 $ & 0.15 & $2\times10^{-4}$ & 0.43 &  0.10 &  $0.06 \pm 0.03$ & 0.06 &	0.001 &	0.44 &	0.16  \\
    \cmidrule{2-12}
    \multirow{4}{*}{{\textbf{citations}}} & \text{Revision Effect} &  $ 2.1 \pm 0.7 $  &  0.22 & $< 10 ^{-6}$ & 0.44 & 0.10  &  $0.10 \pm 0.01$	& 0.10 & $< 10 ^{-6}$	& 0.45 & 0.16 \\
    \cmidrule{2-12}
    & \text{Effect of Tweeting} & $ 2.0 \pm  0.4 $ & 0.18 &  $< 10 ^{-6}$ & 0.56 & 0.14 & $0.05 \pm 0.02$ & 0.04 & $< 10 ^{-6}$ & 0.47 & 0.19  \\
    \cmidrule{2-12}
    & \text{Effect of Author(s) } & $2.6 \pm 2.3$ & 0.20 & 0.02 & 0.58 & 0.12 & $0.20 \pm 0.09$ & 0.16 & $< 10 ^{-6}$ &	0.54 & 0.17 \\
    & \text{Tweeting} & & & & & & & \\
    [1ex]

    \bottomrule
    \end{tabular}}
    \captionof{table}{Average treatment effect (ATE) using standardization of each of our treatments on all citations and highly influential citations over 5 years after first publication. ``Relative ATE'' is calculated by dividing the ATE by the mean outcome value.}
  \label{tab:ate_results_5_years_std}
\end{table*}

\begin{table}[!htb]
    \centering
    \resizebox{1.\columnwidth}{!}{%
    \begin{tabular}{l|l|c|c|c|c}
    \toprule
    
    \multirow{2}{*}{\textbf{Treatment}} & \multirow{2}{*}{\textbf{Year}} & \multicolumn{2}{c}{\bf Computer Science} & \multicolumn{2}{c}{\bf Physics} \\ 
   
   \cline{3-6}
   \\ [0.01ex]
      & & \bf  TMLE ATE & \bf p-value & \bf TMLE ATE & \bf p-value \\ 
  
    \midrule[0.01pt]
    
      &  First & $\mathbf{0.5 \pm 0.5}$ & $\mathbf{0.039}$ & $\mathbf{0.37 \pm 0.23}$  & $\mathbf{0.001}$  \\

    Arxiv-first & Second &  $\mathbf{2.7 \pm 1.7}$ & $\mathbf{0.001}$  &  $\mathbf{0.70 \pm	0.23}$ & $\mathbf{< 10 ^{-6}}$  \\

    effect & Third &  $\mathbf{4.7 \pm  3.6}$ & $\mathbf{0.010}$  & $\mathbf{0.93 \pm	0.27}$ & $\mathbf{< 10 ^{-6}}$  \\

    & Fourth & $6 \pm  6.1$ & $0.051$ &  $\mathbf{0.90 \pm	0.24}$ & $\mathbf{< 10^{-6}}$    \\

    & Fifth &   $\mathbf{7.2 \pm  6}$ & $\mathbf{0.017}$ & $\mathbf{0.74 \pm	0.29}$ & $\mathbf{< 10^{-6}}$  \\
    
    \midrule[0.5pt]
    &  First &  $\mathbf{0.5 \pm  0.4}$ & $\bf{0.040}$ & $\mathbf{0.71	\pm 0.12}$ & $\mathbf{< 10 ^{-6}}$   \\

    Revision & Second &  $\mathbf{2.5 \pm  1.4}$ & $\mathbf{0.008}$ & $\mathbf{0.93 \pm 0.17}$ & $\mathbf{< 10 ^{-6}}$  \\

    effect & Third &  \textbf{4.5 $\pm$  3.3} & $\mathbf{0.008}$ & $\mathbf{1.0 \pm	0.19}$ & $\mathbf{< 10 ^{-6}}$  \\

    & Fourth &  $5.6 \pm 5.7$ & 
    $0.056$ & 
    $\mathbf{0.89 \pm 0.20}$ & $\mathbf{< 10 ^{-6}}$ \\

    & Fifth &  $4.9 \pm  7.4$ & $0.194$ & 
    $\mathbf{0.79 \pm 0.20}$ & $\mathbf{< 10 ^{-6}}$ \\
    
    \midrule[0.5pt]
     &  First &  $\mathbf{1.5 \pm  0.2}$ & $\mathbf{< 10 ^{-6}}$ & 
     $\mathbf{1.65 \pm 1.03}$ & $\mathbf{0.002}$  \\

    Effect of & Second &  $\mathbf{4.9 \pm 0.7}$ & $\mathbf{< 10 ^{-6}}$ & $\mathbf{1.08 \pm 0.63}$ & $\mathbf{7 \times 10 ^{-4}}$ \\

    tweeting & Third & $\mathbf{9.9 \pm  1.6}$ & $\mathbf{< 10 ^{-6}}$ & $\mathbf{1.90 \pm 0.79}$ & $\mathbf{< 10 ^{-6}}$ \\

    & Fourth & $\mathbf{13.9 \pm  2.5}$ & $\mathbf{< 10 ^{-6}}$ & $\mathbf{1.52	\pm 0.64}$ & $\mathbf{< 10 ^{-6}}$ \\

    & Fifth &  $\mathbf{14.0 \pm  3.2}$ & $\mathbf{< 10 ^{-6}}$  & $\mathbf{1.54	\pm 0.57}$ & $\mathbf{< 10 ^{-6}}$ \\
    \midrule[0.5pt]
    &  First &  $\mathbf{1.1 \pm  0.3}$ & $\mathbf{< 10 ^{-6}}$  & $\mathbf{0.37 \pm 0.10}$ & $\mathbf{< 10 ^{-6}}$ \\

   Effect of & Second &  $\mathbf{ 1.1 \pm  1}$ & $\mathbf{0.037}$ & ${-0.06 \pm 0.62}$ & ${0.854}$ \\

    author(s)  & Third & $1.9 \pm 2.03$ & $0.072$  & $\mathbf{0.55	\pm 0.13}$ & $\mathbf{< 10^{-6}}$ \\

     tweeting & Fourth &  $\mathbf{4.3 \pm  3.2}$ & $\mathbf{0.008}$ & ${0.36 \pm 1.12}$ & $0.53$   \\

    & Fifth &   $\mathbf{13.5 \pm  5.5}$ & $\mathbf{< 10 ^{-6}}$ & ${0.14 \pm 0.39}$ & $0.48$   \\
    
    \bottomrule
    \end{tabular}%
    }
    \vspace{0.7cm}
    \caption{Average treatment effects for all citations using each treatment variable for the first through fifth year of publication. Statistically significant results ($p<0.05$) are in bold.}
    \label{tab:ate_over_years}
\end{table}


\begin{table}[!htb]
    \centering
    \resizebox{1.\columnwidth}{!}{%
    \begin{tabular}{l|l|c|c|c|c}
    \toprule
    
    \multirow{2}{*}{\textbf{Treatment}} & \multirow{2}{*}{\textbf{Year}} & \multicolumn{2}{c}{\bf Computer Science} & \multicolumn{2}{c}{\bf Physics} \\ 
   
   \cline{3-6}
   \\ [0.01ex]
      & & \bf  TMLE ATE & \bf p-value & \bf TMLE ATE & \bf p-value \\ 
  
    \midrule[0.01pt]
    
      &  First & $0.09 \pm 0.10$ & $0.108$ & ${0.03 \pm 0.03}$  & ${0.06}$  \\

    Arxiv-first & Second &  $\mathbf{0.49 \pm 0.36}$ & $\mathbf{0.008}$  &  $\mathbf{0.07 \pm 0.03}$ & $\mathbf{< 10 ^{-6}}$  \\

    effect & Third &  $0.59 \pm  0.90$ & $0.196$  & $\mathbf{0.08 \pm 0.02}$ & $\mathbf{< 10 ^{-6}}$  \\

    & Fourth & $\mathbf{0.98 \pm  0.96}$ & $\mathbf{0.046}$ &  $\mathbf{0.06 \pm 0.03}$ & $\mathbf{2.5 \times 10^{-5}}$    \\

    & Fifth &   $\mathbf{1.10 \pm  0.88}$ & $\mathbf{0.015}$ & $\mathbf{0.05 \pm 0.03}$ & $\mathbf{0.007}$  \\
    \midrule[0.5pt]
    &  First &  $0.04 \pm  0.10$ & $0.426$ & $\mathbf{0.03 \pm 0.012}$ & $\mathbf{< 10 ^{-6}}$   \\

    Revision & Second &  $\mathbf{0.46 \pm  0.33}$ & $\mathbf{0.005}$ & $\mathbf{0.05 \pm 0.02}$ & $\mathbf{< 10 ^{-6}}$  \\

    effect & Third &  0.73 $\pm$  78 & $0.063$ & $\mathbf{0.05 \pm 0.02}$ & $\mathbf{< 10 ^{-6}}$  \\

    & Fourth &  $0.76 \pm 1.28$ & $0.244$ & $\mathbf{0.04 \pm 0.02}$ & $\mathbf{< 10 ^{-6}}$ \\

    & Fifth &  $0.55 \pm  1.60$ & $0.501$ & $\mathbf{0.03 \pm 0.02}$ & $\mathbf{5 \times 10 ^{-5}}$ \\
    \midrule[0.5pt]
     &  First &  $\mathbf{0.32 \pm  0.05}$ & $\mathbf{< 10 ^{-6}}$ & $\mathbf{0.09 \pm 0.09}$ & $\mathbf{ 0.05}$  \\

    Effect of & Second &  $\mathbf{1.01 \pm 0.16}$ & $\mathbf{< 10 ^{-6}}$ & $\mathbf{0.08 \pm 0.05}$ & $\mathbf{0.002}$ \\

    tweeting & Third & $\mathbf{1.79 \pm  0.33}$ & $\mathbf{< 10 ^{-6}}$ & $\mathbf{0.09 \pm 0.05}$ & $\mathbf{7 \times 10 ^{-5}}$ \\

    & Fourth & $\mathbf{2.20 \pm  0.50}$ & $\mathbf{< 10 ^{-6}}$ & $\mathbf{0.06 \pm 0.06}$ & $\mathbf{0.05}$ \\

    & Fifth &  $\mathbf{2.11 \pm  0.61}$ & $\mathbf{< 10 ^{-6}}$  & $\mathbf{0.09 \pm 0.06}$ & $\mathbf{0.001}$ \\
    \midrule[0.5pt]
    &  First &  $\mathbf{0.18 \pm  0.05}$ & $\mathbf{3.2 \times 10 ^{-6}}$  & $\mathbf{0.07 \pm 0.02}$ & $\mathbf{< 10 ^{-6}}$ \\

   Effect of & Second &  $\mathbf{ 0.63 \pm  0.27}$ & $\mathbf{6.8 \times 10^{-6}}$ & $\mathbf{0.03 \pm 0.01}$ & $\mathbf{< 10 ^{-6}}$ \\

    author(s)  & Third & $0.31 \pm 0.47$ & $0.187$  & ${0.01 \pm 0.03}$ & ${0.06}$ \\

     tweeting & Fourth &  $\mathbf{2.43 \pm  1.15}$ & $\mathbf{3.2 \times 10^{-5}}$ & $\mathbf{0.10 \pm 0.04}$ & $\mathbf{< 10 ^{-6}}$   \\

    & Fifth &   $\mathbf{1.74 \pm  0.80}$ & $\mathbf{2 \times 10 ^{-5}}$ & $\mathbf{0.01 \pm 0.01}$ & $\mathbf{0.03}$   \\
    
    \bottomrule
    \end{tabular}%
    }
    \vspace{0.7cm}
    \caption{Average treatment effects for highly influential citations using each treatment variable for the first through fifth year of publication. Statistically significant results ($p<0.05$) are in bold.}
    \label{tab:ate_over_years_highly_inf_cit}
\end{table}

\subsection{E. Average treatment effects using standardization}

We find statistically significant positive effects for all our treatments for all citations and highly influential citations across computer science and physics. The  results are listed in Table \ref{tab:ate_results_5_years_std}.

\subsection{F. Average treatment effects for the first through fifth year of publication}

Table \ref{tab:ate_over_years} and  Table \ref{tab:ate_over_years_highly_inf_cit} provide details about the treatment effects over the years for all citations and highly influential citations, respectively.

\end{document}